\documentclass[reprint,aps,prd,superscriptaddress,preprintnumbers,nofootinbib]{revtex4-1}

\usepackage{mathrsfs} 
\usepackage{amsfonts}
\usepackage{amsmath}
\usepackage{amsthm}
\usepackage{amssymb}
\usepackage{array}
\usepackage{dcolumn}
\usepackage[normalem]{ulem}
\usepackage[svgnames]{xcolor}
\usepackage[hidelinks]{hyperref}
\hypersetup{
    colorlinks=true,
    linkcolor=NavyBlue,
    filecolor=NavyBlue,
    urlcolor=NavyBlue,
    citecolor=NavyBlue,
    }

\usepackage{graphics}
\usepackage{graphicx}
\usepackage[caption=false]{subfig}
\usepackage{adjustbox}
\usepackage{breqn}
\usepackage{fontawesome} 
\usepackage{slashed}

\usepackage{tipa}

\def\be{\begin{equation}}
\def\ee{\end{equation}}
\def\bea{\begin{eqnarray}}
\def\eea{\end{eqnarray}}

\newcount\Comments  
\newcommand{\kibitz}[2]{\ifnum\Comments=1\textcolor{#1}{#2}\fi}

\makeatletter
\let\cat@comma@active\@empty
\makeatother

\begin{document}

\title{Multi-Messenger Astrology}

\author{Gwen Walker}
\affiliation{Department of Physics, Virginia Tech, Blacksburg, Virginia 24061, USA}
\affiliation{Scorpio}

\author{Nick Ekanger}
\affiliation{Department of Physics, Virginia Tech, Blacksburg, Virginia 24061, USA}
\affiliation{Taurus}

\author{R. Andrew Gustafson}
\affiliation{Department of Physics, Virginia Tech, Blacksburg, Virginia 24061, USA}
\affiliation{Pisces}

\author{Sean Heston}
\affiliation{Department of Physics, Virginia Tech, Blacksburg, Virginia 24061, USA}
\affiliation{Libra}

\begin{abstract}
{{It has long been accepted that the cosmos determine our personalities, relationships, and even our fate. Unlike our condensed matter colleagues - who regularly use quantum mechanics to determine the healing properties of crystals - astrology techniques have been unchanged since the 19th century. In this paper, we discuss how astrophysical messengers beyond starlight can be used to predict the future and excuse an $\mathcal{O}(1)$ fraction of our negative personality traits.}  \\}

\end{abstract}

\maketitle

\section{Introduction} 

Since the dawn of time\footnote{c. 1981}, man has looked to the stars. After a brief period of pure admiration, he started to use these mythical light sources to answer life's most important questions. Astronomers stick to relatively boring questions: \textit{``Where did we come from?", ``What are we made of?", ``Where are we going?"}. Meanwhile, astrologers such as ourselves focus on using the cosmos for more important questions: \textit{``How can we impress people at the bar?"}, and \textit{``How can I make the breakup not my fault?"}.

However, our nerdy astronomy peers have recently made some worthwhile progress.\footnote{Even a stopped clock is right twice a day.} They have begun looking into the sky not just for visible wavelengths of light, but for other particles as well. These searches have been named ``multi-messenger astronomy'', a rather ironic name considering how rarely astronomers get messages. Still, this shows a gap in our understanding as astrologers. For years we have been studying the motion of the planets through the visible constellations, speaking of Mercury in Gatorade, but we never considered Mercury in gamma-ray Gatorade (henceforth referred to as Gammarade).

In this paper, we seek to explain how these invisible yet powerful astrophysical sources influence our destiny, personalities, and ultimately fruitless love-lives. We start with astronomers' understanding of these messengers, and provide the definitive astrological determinations for each.\footnote{Subject to change if it helps us talk up the barista.} However, we must be realistic, as this paper needs to go out on April 1st, and it's mid-March now, so we offer two important caveats. First, we do not offer a comprehensive guide on how each sign changes under the multi-messenger frame work, and leave it as an exercise to the reader to use the available data and avoid personal responsibility. Second, we make no mention of astrological ``houses'', as we are poor and can only afford apartments.

\begin{figure*}
    \centering
    \includegraphics[width=0.45\linewidth]{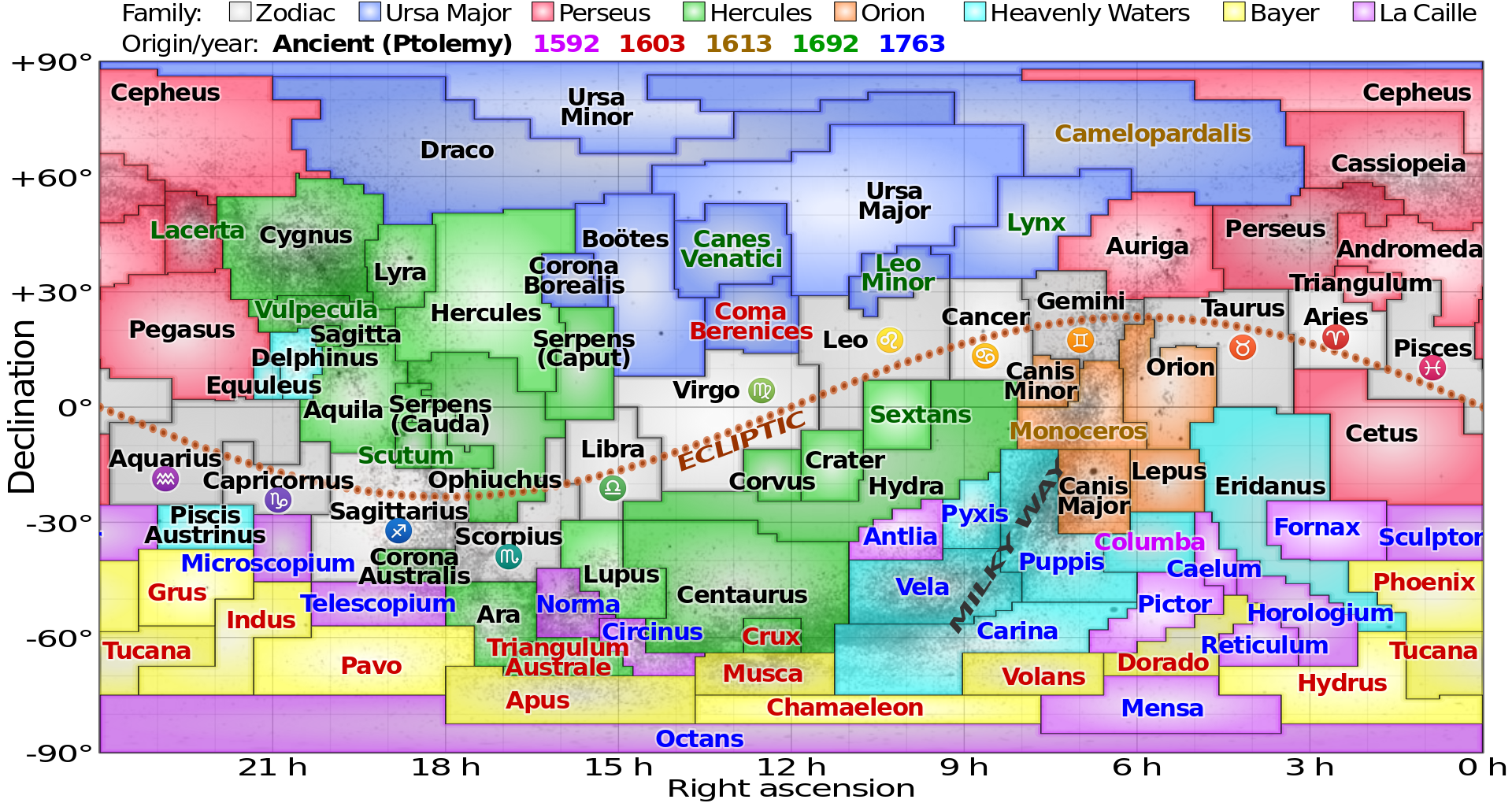}
    \includegraphics[width=0.45\linewidth]{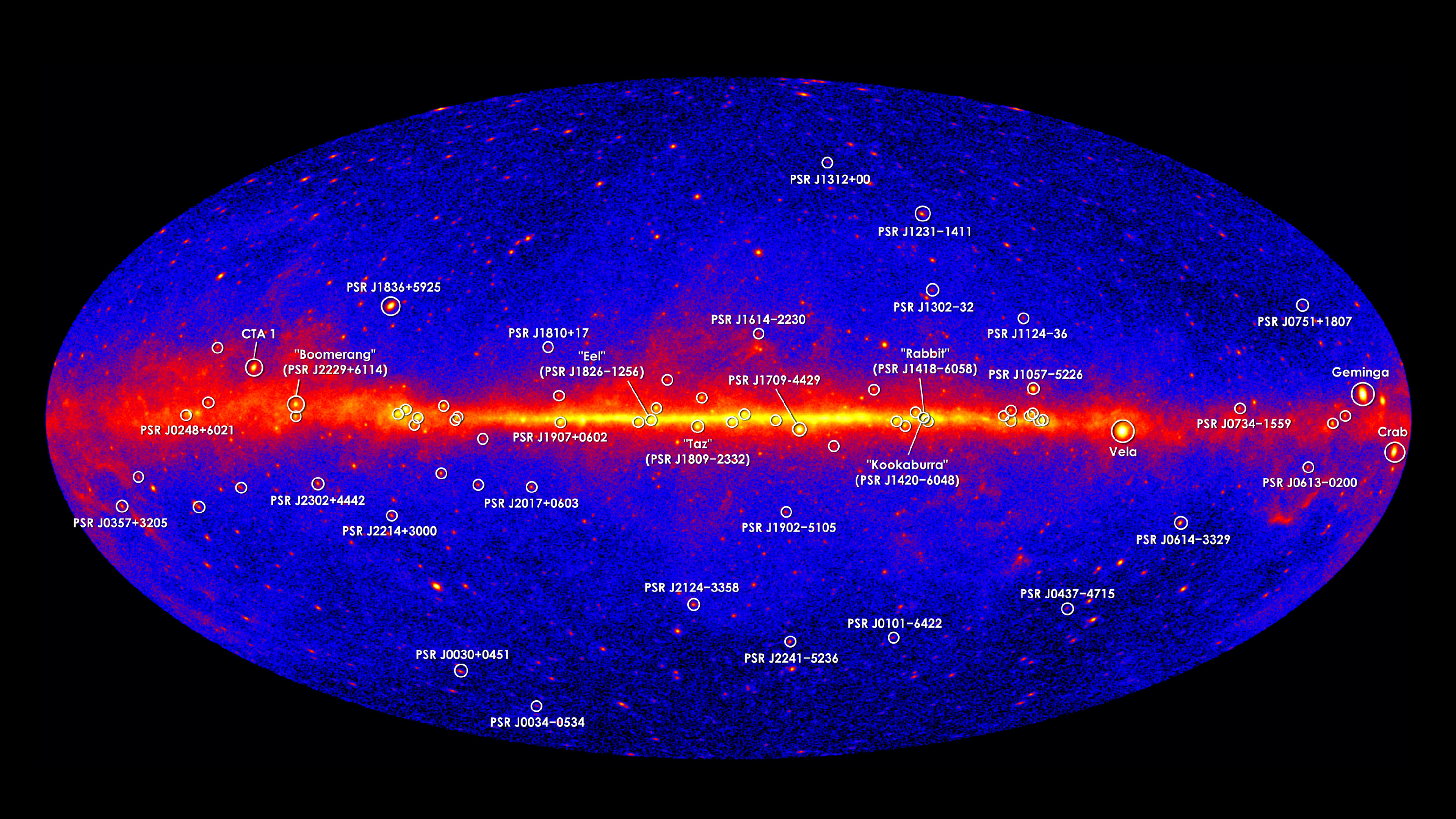}
    \\
    \includegraphics[width=0.45\linewidth]{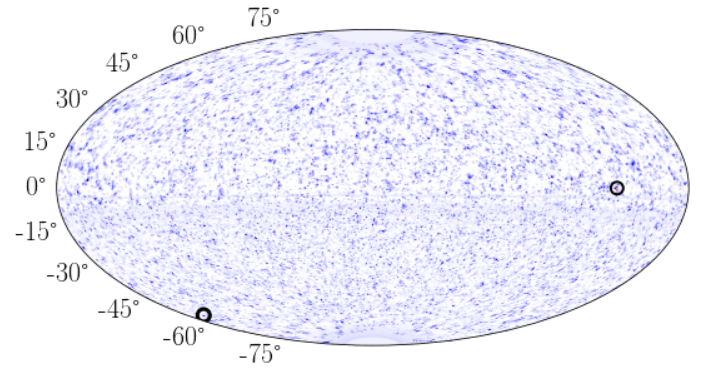}
    \includegraphics[width=0.45\linewidth]{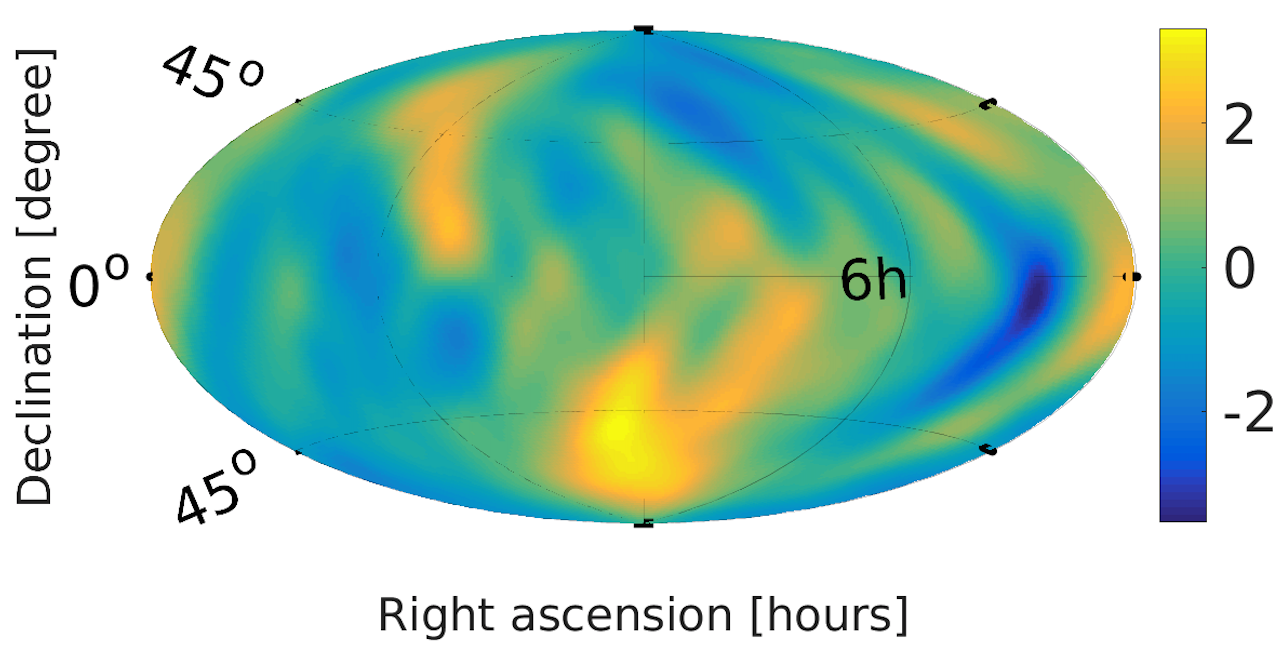}
    \caption{Comparison of the sky through a multi-messenger lens for starlight \cite{WikipediaAstrologicalSign} (\textit{upper left}), gamma-rays \cite{GammaRaySky} (\textit{upper right}), neutrinos \cite{NeutrinoSky} (\textit{lower left}), and gravitational waves \cite{GWSky} (\textit{lower right}). The cosmic ray constellation map will not be included because it is outside the scope of this paper (i.e. it is too much work and I don't want to). Although the starlight map clearly creates signs that control our personalities and is well understood, more data is needed (c'mon, just a little more) in all other messengers to determine \textit{(i)} if there are Zodiac signs in gamma-ray, neutrino, gravitational wave, and cosmic ray sky (which there must be to explain our sometimes erratic behavior) but also \textit{(ii)} if they correlate with the traditional Zodiac signs.}
    \label{fig:constellationmaps}
\end{figure*}

\section{Multi-Messenger Signals}

In this section, we cover the effects of different messengers on one's personality. We go over the known messengers that we have observed in the past either as persistent sources or transients. 

\subsection{Gamma-rays}
High exposure to gamma-rays (\textipa{\textdyoghlig \ae m.\textrhookschwa}- \textipa{\textturnr e\i z}) is well accepted to cause anger management issues (see, e.g., The Hulk). Therefore, if there are strong persistent gamma-ray sources, like the excess gamma-rays from the center of the Milky Way, existing within your constellation, you are more prone to lash out in anger. There are other special circumstances for people born at a time when high energy photons from gamma-ray bursts (GRBs) arrive at Earth. However, how strong of an influence these gamma-rays from GRBs can have on your personality is highly dependent upon the flux, and therefore the distance, to the source. This should therefore be treated on a case-by-case basis. Luckily, the GCN circulars page exists (\url{https://gcn.nasa.gov/circulars}), and one can easily search for gamma-ray events filtered by date if one's day of birth is after 1997. For those older than this, there is no well-known record of gamma-ray events, so you are on your own to find out if your personality has been affected by excess gamma-ray exposure. 

\subsection{Cosmic Rays}

Cosmic rays are charged particles which have been accelerated to high energies. To help quantify these energies, we introduce a new unit called a Baseball Speed (BS), which is the average energy of a thrown baseball (about 50\,J). Despite having the name ``speed'' in the name, a BS is a unit of energy. We apologize for any confusion that this causes, especially because astronomy is known for its clear and intuitive units of measurement. The majority of the cosmic-ray flux has energies in the fBS-pBS range.

\begin{figure}
    \centering
    \includegraphics[width = 0.9\linewidth]{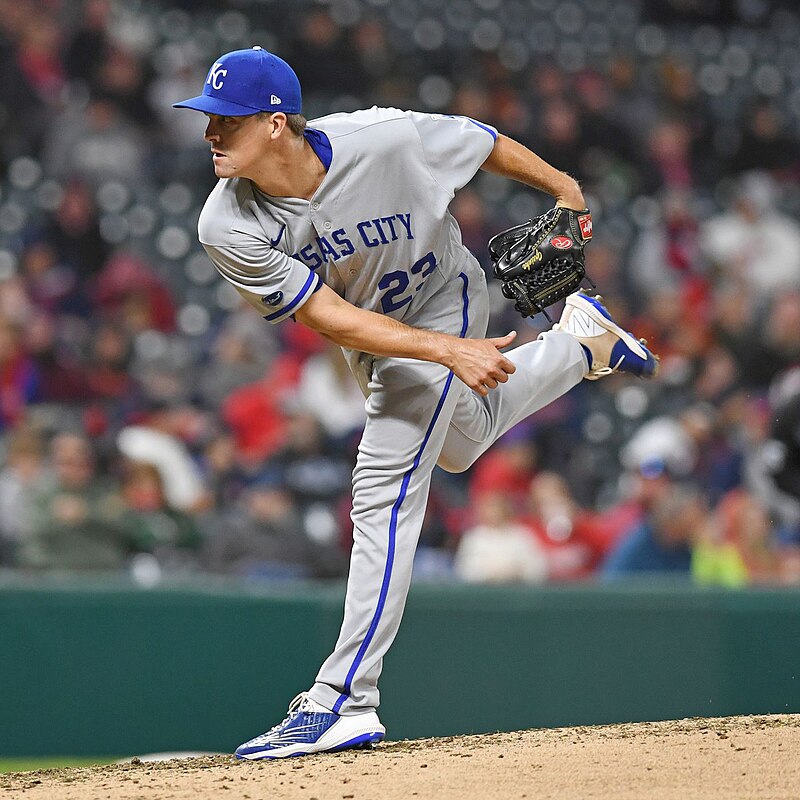}
    \caption{An accelerator of (technically) astrophysical origin, capable of accelerating objects up to 1 BS \cite{Baseball}.}
    \label{bs}
\end{figure}

Cosmic rays are the only messenger with electric charge, and they therefore dictate your ``electronic personality'': the personality you project through text, social media, and email for those older than z = $3\times 10^{-9}$. This electronic personality does not necessarily match the personality dictated by other messengers. For example, when the authors of this paper meet online friends in person, these friends are often upset that we are not ``fun to be around'' or ``six-foot-two'' like our profiles proclaim. Although this was previously attributed to lying and insecurity, we can now see that it stems from a difference in our cosmic ray astrological signs and our visible astrological signs.

While steady sources of cosmic rays determine our average electronic personality, short-lived high energy bursts of cosmic rays can cause unusual online behavior. Last Tuesday, high-energy cosmic rays on the order of $\mu$BS-mBS are responsible for us buying a four-wheeler online and texting our exes (we don't have a source for this, but it's the only thing that makes sense).

\subsection{Neutrinos}
Neutrinos are often referred to as ghost particles since, like ghosts, they rarely interact and are produced in nuclear reactors. The weak interactions mean that neutrinos carry pointing information that would be lost by gamma-ray/cosmic-ray scattering. In an astrological sense, this ghostly nature means that those born under neutrinos have a higher sense of spirituality and, more importantly, are more likely to ghost\footnote{``Ghosting'' is a colloquial term for not responding to messages. It's often seen as incredibly rude, but ultimately fine when I do it because others are harshing my mellow.} others.

A major source of neutrinos is the Sun with a myriad of fusion processes which power the core. For an example of this effect, look no further than the authors of this paper. When the Sun rises, the influx of neutrinos cause us to ghost our dates from the previous night. This explanation is favored over the previous hypothesis of ``commitment issues'' by $2.7\sigma$.

Another astrological source of neutrinos are core-collapse supernovae. The last such supernova close enough to produce observable neutrinos was in 1987, with an enormous flux at Earth. This affected many people at the time, although the effects may lie dormant for years. This is seen in the case our fathers, who took until the early 2000s to ghost their respective families.

In recent years, IceCube (the experiment, not the rapper) has identified point-sources for high energy neutrinos. This work is still recent, but in time it will allow astrologers to match up which signs have the most neutrinos and are most likely to ghost (but let's be real, we know it's going to be Virgo).

\subsection{Gravitational Waves}
Einstein's theory of General Relativity predicts waves of contracting and expanding spacetime propagating at the speed of light. Einstein also married two of his cousins, so he was clearly focused on the wrong type of attraction and relativity.

Gravitational waves are the youngest of the astrological messengers, but that does not make them any less loved by astronomers. They are less loved by astrologers, though, because the primary observatory is based out of Louisiana, and humidity is murder on our hair. Gravity is a force which is always attractive, and gravitational waves are formed when two objects merge. Therefore, in an astrological framework, gravitational waves dictate your attractiveness and your attraction to others. We think that's obvious; it's honestly a little embarassing that you needed us to explain it to you.

\section{Example Multi-Messenger Astrological Reading}

Take, for example, an Aries. In Figure~\ref{fig:constellationmaps}, a trained multi-messenger astrologer can see that this sign also lies within the gamma-ray signal, a persistent neutrino source, and a potential gravitational wave void. The common perception of an Aries is that they are natural go-getters that are fair and patient in relationships \cite{AriesUSATODAY}. Although they may be strong leaders, a more nuanced reading reveals their relationships may be more shaky as their position in the sky is coincident with ire-inducing gamma-rays and relationship-rocking neutrinos. The role of gravitational waves on personalities is still being quantified, but given that they are in a void, they probably won't marry their cousins.

\section{Future Possible Searches}

\subsection{Dark Matter}
Dark matter is a currently undiscovered type of exotic matter, proposed to explain galaxy rotation curves, source micro-lensing events, fix marriages, and whatever else the phenomenologists want to solve that week. Therefore, it is difficult to extract how dark matter detection could affect one's personality. One trait for individuals with directional dark matter detected within their sign is that they should only rarely interact with others. 

\subsection{ALPs}
Axion-like particles (ALPs), are proposed pseudoscalar bosons, popularized by the 1980's TV sitcom of the same name. Although this particle was originally theorized to solve the Strong CP problem and get into wacky weekly hijinks, it has been shown that if they exist, then they must be produced in different astrophysical scenarios.

\begin{figure}
    \centering
    \includegraphics[width = 0.9 \linewidth]{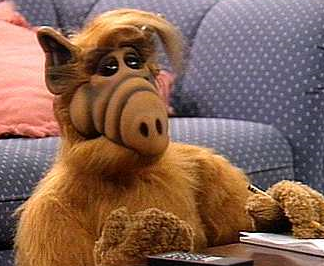}
    \caption{Image of a theoretical ALP from 3 years after the Dawn of Time \cite{Wikipedia}.}
    
\end{figure}

Many of the astrological properties of an ALP flux are model dependent, and beyond the scope of this paper. However, all models include the pseudoscalar nature of these lovable-goofball particles, so from this we have determined that being born under a flux of ALPs will make a person more susceptible to pseudoscience. Thankfully, none of the authors of this paper have such a problem.

\section{Summary and Conclusion}
While this paper is not intended to describe the astrological properties of messengers in an exhaustive way, it has left us exhausted, so we leave further analyses to smarter, harder-working individuals. In the meantime, the larger astrological society should be cognizant of the fact that destiny and personality is determined by more than just visual photons. Even with our limited, unfunded\footnote{Can you believe that everyone rejected our proposal?} work, we have been able to better explain away many of our personal ``flaws" in this new framework. We highly encourage our colleagues to do the same.

\section*{Acknowledgements}
We would like to thank all of our coworkers who told us this was a bad idea, and the readers of this paper for having a sense of humor. We especially thank our future employers for not holding this against us.

\bibliographystyle{apsrev4-1.bst}
\bibliography{main.bib}

\end{document}